\documentclass[reprint,superscriptaddress,amsmath,amssymb,aps,prl,floatfix,
]{revtex4-2}

\usepackage{graphicx}
\usepackage{dcolumn}
\usepackage{bm}

\begin{document}
\title{Sub-recoil clock-transition laser cooling enabling shallow optical lattice clocks}

\author{X. Zhang}
\affiliation{National Institute of Standards and Technology, 325 Broadway, Boulder, Colorado 80305, USA}
\affiliation{University of Colorado, Department of Physics, Boulder, Colorado 80309, USA}

\author{K. Beloy}
\affiliation{National Institute of Standards and Technology, 325 Broadway, Boulder, Colorado 80305, USA}

\author{Y. S. Hassan}
\affiliation{National Institute of Standards and Technology, 325 Broadway, Boulder, Colorado 80305, USA}
\affiliation{University of Colorado, Department of Physics, Boulder, Colorado 80309, USA}

\author{W. F. McGrew}
\altaffiliation{Present address: JILA, University of Colorado and National Institute of Standards and Technology, Boulder, Colorado 80309, USA.}
\affiliation{National Institute of Standards and Technology, 325 Broadway, Boulder, Colorado 80305, USA}
\affiliation{University of Colorado, Department of Physics, Boulder, Colorado 80309, USA}

\author{C.-C. Chen}
\author{J. L. Siegel}
\author{T. Grogan}
\affiliation{National Institute of Standards and Technology, 325 Broadway, Boulder, Colorado 80305, USA}
\affiliation{University of Colorado, Department of Physics, Boulder, Colorado 80309, USA}

\author{A. D. Ludlow}
\email{andrew.ludlow@nist.gov}
\affiliation{National Institute of Standards and Technology, 325 Broadway, Boulder, Colorado 80305, USA}
\affiliation{University of Colorado, Department of Physics, Boulder, Colorado 80309, USA}

\date{\today}
\begin{abstract}
  Laser cooling is a key ingredient for quantum control of atomic systems in a variety of settings. In divalent atoms, two-stage Doppler cooling is typically used to bring atoms to the {\textmu}K regime. Here, we implement a pulsed radial cooling scheme using the ultranarrow $^{1}$S$_{0}$-$^{3}$P$_{0}$ clock transition in ytterbium to realize sub-recoil temperatures, down to tens of nK. Together with sideband cooling along the one-dimensional lattice axis, we efficiently prepare atoms in shallow lattices at an energy of 6 lattice recoils. Under these conditions key limits on lattice clock accuracy and instability are reduced, opening the door to dramatic improvements. Furthermore, tunneling shifts in the shallow lattice do not compromise clock accuracy at the 10$^{-19}$ level.
\end{abstract}

\maketitle
Laser cooled and trapped ionic, atomic, and molecular systems have realized exceptional quantum control. As a result, these systems have been ideal for fundamental physics studies \cite{SM2018}, explorations of many-body physics \cite{MB2008, MBm2013}, quantum computation implementation \cite{Qc1999, Qc2002}, quantum information applications \cite{Qi2017, Qim2017}, and precision measurements \cite{OC2015, Pm2017}. Over recent decades, atoms with two valence electrons have attracted significant attention for their enhanced capability of quantum control. Notably, these atoms possess both a ground and metastable excited state with zero electronic angular momentum, offering quantum coherence on timescales of seconds or beyond.

These atomic structure features are prominently exploited in optical lattice clocks. Remarkable quantum coherence has been experimentally realized using the `clock' transition between these states, enabling unprecedented levels of frequency accuracy at the 10$^{-18}$ fractional level \cite{Yb2018, Sr2019}. As a result, these clocks can be used as a redefining anchor for the International System of Units \cite{Rs2018}, to test the variation of fundamental constants \cite{Fc2021}, to measure Earth's geopotential \cite{Geo2018, Geo2020}, and to search for dark matter \cite{Dm2018, Dm2020, Dm2021}. But as the performance in these applications improves, new considerations require even greater levels of quantum control. This control begins with lower atomic temperatures to minimize trapping inhomogeneity and maximize coherence. 

For divalent atoms like ytterbium, strontium, and mercury, Doppler cooling on the narrow intercombination $^{1}$S$_{0}$-$^{3}$P$_{1}$ transition typically affords atomic temperatures in the range of one to tens of {\textmu}K \cite{YbMOT2009, SrPTB2014, SrLattice2018, Sr2019, YbInrim2020}. Evaporative cooling has been used to attain sub-{\textmu}K temperature and quantum degeneracy \cite{3Dclock2017, QS2020}, but is precluded in many applications because of atom loss and very long evaporation time. Techniques for additional cooling of strongly-confined atoms have recently been demonstrated, but these are generally limited to the resolved-sideband regime \cite{Lattice2017, Lattice2016, SrLattice2018, TB2021, Tweezer2021}. Many trapped systems operate outside this regime, including most optical lattice clocks that employ a 1D lattice for metrological benefits.


Here we realize greater levels of quantum control with sub-recoil laser cooling on the doubly forbidden clock transition (natural linewidth $\approx$ 8 mHz) in ytterbium, reaching temperatures down to tens of nK. In so doing, we help to resolve several critical problems in optical lattice clocks. While confining atoms in a `magic wavelength' optical trap helps reject lowest-order light shifts on the clock transition, higher-order effects make complete elimination impossible \cite{Lattice2017, SrLattice2018, YbLattice2019}. Moreover, Raman scattering of optical lattice photons quenches the excited metastable state, limiting the desired quantum coherence \cite{Srscatter2018, Srscatter2019}. Here, the pulsed clock-transition cooling enables efficient loading of shallow lattices down to 6 E$_{\text{r}}$ $\approx$ k$_{\text{B}} \times$ 600 nK (E$_{\text{r}}$ = $h^{2}/2m\lambda_{l}^{2}$ is the lattice recoil energy, $\lambda_{l}$ is the optical lattice wavelength, $h$ is Planck's constant, m is the atomic mass of $^{171}$Yb isotope, and k$_{\text{B}}$ is the Boltzmann constant).  At these depths, lattice light shifts and $^{3}$P$_{0}$ excited state quenching are strongly suppressed, making systematic uncertainty at the $10^{-19}$ level or below feasible. The nK-regime temperatures enable resolving motional transitions from distinct lattice bands, allowing extra control. Strong intersite tunneling is observed in the form of Bloch oscillations at 6 E$_{\text{r}}$, inducing a minimal frequency shift. The pulsed clock-transition cooling demonstrated here can benefit neutral atom quantum computing architectures \cite{QI2011, QI2019, QS2020}, where lower temperatures suppress thermal dephasing to improve entanglement fidelity and qubit control \cite{QI2020, Tweezer2021} or to enhance single-atom detection fidelity \cite{QI2019S}. It also benefits strategies for direct cooling towards quantum degeneracy \cite{LCQD2013, Rb2017, Rb2019}, as well as simulations of quantum magnetism, Kondo lattice physics \cite{Kondo2010}, and other Hamiltonians \cite{QI2011, QS2020}.

\begin{figure}[htbp]
\includegraphics{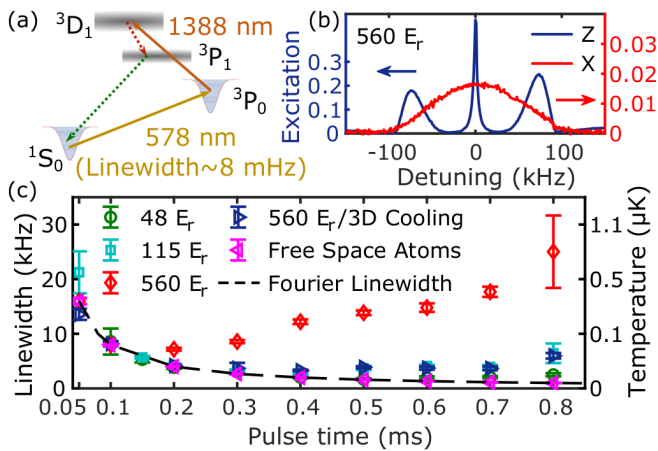}
\caption{\label{fig:1} (a) Energy levels used in the cooling. Note that spontaneous decay from $^{3}$D$_{1}$ (lifetime $\sim$330 ns) follows the branching ratios 0.64 ($^{3}$P$_{0}$), 0.35 ($^{3}$P$_{1}$), and 0.01 ($^{3}$P$_{2}$, not shown). $^{3}$P$_{1}$ population decays to the ground state with a $\sim$870 ns lifetime. (b) Atomic excitation as a function of laser detuning from the 578-nm transition for a laser propagating along the \emph{x} (red) and \emph{z} (blue) axes at 560 E$_{\text{r}}$ lattice depth. (c) Measured linewidth of velocity selection profiles as a function of the first 578-nm laser pulse duration and for different trap depths [48 E$_{\text{r}}$(green circle), 115 E$_{\text{r}}$(cyan square), and 560 E$_{\text{r}}$(red diamond)] and cooling conditions [560 E$_{\text{r}}$ after three-dimensional (3D) cooling (blue triangle)]. The magenta triangle data are measured with free-space atoms and the black dashed line is the calculated Fourier-limited linewidth for different pulse durations. The right ordinate indicates the corresponding temperature of the selected atoms, from free-space Doppler theory.}
\end{figure}


Many details of our experimental apparatus are described elsewhere \cite{Yb2018}. After Doppler cooling on the $^{1}$S$_{0}$-$^{1}$P$_{1}$ and $^{1}$S$_{0}$-$^{3}$P$_{1}$ transitions, atoms are loaded into a `magic wavelength' one-dimensional (1D) optical lattice at 759 nm. The lattice is formed using a power enhancement cavity with a 1/e field radius of 170 {\textmu}m and aligned with $\leq$1$^{\text{o}}$ offset from gravity. We label this longitudinal axis \emph{z}, and the radial axes \emph{x} and \emph{y}. One pair of counter-propagating, orthogonally polarized 578-nm laser beams (waist = 400 {\textmu}m) travels along \emph{x}, while a similar pair also travels along \emph{y}. Both are used for selection in the pulsed clock-transition cooling. Another 578-nm laser propagates along $-z$ for longitudinal sideband cooling, while a final 578-nm laser propagates along $+z$ for narrow-line spectroscopy. A 1388-nm laser, resonant with the $^{3}$P$_{0}$-$^{3}$D$_{1}$ transition (Fig.~1(a)), travels with a small tilt relative to \emph{z} and is used for both sideband cooling and pulsed radial cooling.


Atomic confinement along the radial and longitudinal axes differs significantly in our 1D lattice (strong confinement along \emph{z} and weak confinement along \emph{x, y}). To highlight this, Figure~1(b) shows atomic excitation as a function of laser detuning from the $^{1}$S$_{0}$-$^{3}$P$_{0}$ transition at 578 nm for a laser propagating along the \emph{x} or \emph{z} axes. Along the \emph{z} axis, strong confinement enables well-resolved (albeit motionally broadened) sidebands at red and blue detuning. For spectroscopy along \emph{x}, weak confinement yields an excitation spectrum resembling the familiar Doppler-broadened profile, similar to the case of free-space atoms. The Doppler width is 118.4(12) kHz corresponding to a radial temperature of 17.4(4) {\textmu}K.


Cooling along the radial axes begins by selectively exciting a velocity group within the Doppler-broadened distribution of Fig.~1(b) on the $^{1}$S$_{0}$-$^{3}$P$_{0}$ clock transition. We select two velocity groups by tailoring the counter-propagating 578-nm clock laser intensity, duration, and frequency detuning of the excitation pulse. Afterwards, a pulse of 1388-nm laser light further excites the velocity-selected atoms in $^{3}$P$_{0}$ to $^{3}$D$_{1}$, where they spontaneously decay to $^{1}$S$_{0}$ via $^{3}$P$_{1}$ and the atomic velocity redistributes irreversibly through random recoil kicks \cite{Supp}. By linking together a sequence of 578-nm and 1388-nm laser pulses, the atomic population accumulates in the zero-velocity dark state, which is off-resonance relative to the 578-nm velocity selection laser frequency..

While the cooling principle is related to pulsed Raman cooling \cite{Raman1992, Raman1994}, here we exploit the ultranarrow clock transition for precise single-photon velocity selection, rather than a two-photon Raman process. The coherence of the long-lived $^{3}$P$_{0}$ state facilitates sub-recoil temperatures.  A related technique has also been used to 1D cool $^{40}$Ca atoms in free space using the intercombination $^{1}$S$_{0}$-$^{3}$P$_{1}$ transition \cite{Ca1, Ca2, Ca3}.  Here, atomic confinement not only enables repetitive and long-duration cooling pulses without atom escape, but magic wavelength operation also prevents inhomogeneous trap light shifts from degrading the velocity selection.


However, the trap does introduce a challenge to the velocity selection process. Oscillatory atomic motion along the weak trap axes yields a periodic Doppler shift on the clock transition, broadening the velocity selection profile. To explore this effect experimentally, we excite trapped atoms near zero velocity with a resonant 578-nm laser $\pi$ pulse along \emph{x}. After removing any residual ground state atoms via repeated cycling on the $^{1}$S$_{0}$-$^{1}$P$_{1}$ transition at 399 nm, we use a second, longer-duration 578-nm $\pi$ pulse along \emph{x} with variable detuning to de-excite selected $^{3}$P$_{0}$ atoms back to the ground state, where 399-nm laser fluorescence measurements are made. To ensure that lattice confinement does not impact the de-excitation process, the lattice is abruptly extinguished prior to this second 578-nm pulse. The velocity selection profile is read out as detuning of the second pulse is scanned (see SM Fig.~2 (b) and (c), \cite{Supp}).  


\begin{figure}[htbp]
\includegraphics{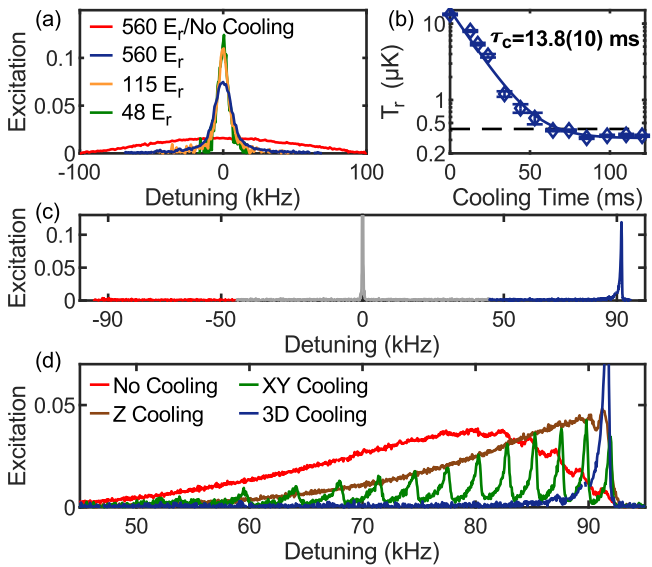}
\caption{\label{fig:2} (a) Radial spectra before (red) and after (blue) pulsed radial cooling at 560 E$_{\text{r}}$. The same cooling process was also realized at 48 E$_{\text{r}}$ (green) and 115 E$_{\text{r}}$ (orange). The corresponding radial temperatures of atoms are 17.4(4) {\textmu}K, 250(10) nK, 90(10) nK, and 90(10) nK, respectively. The total cooling time for each trap depth is 84 ms, 93 ms, and 96 ms, respectively. Temperature is derived from the width of a Voigt line shape fit \cite{Supp}. (b) Measured radial temperature as a function of cooling time at 560 E$_{\text{r}}$. The cooling time is adjusted by varying the number of pulse cycles applied. The solid line is the exponential fit. The black dashed line is the recoil-limited temperature of 410 nK. (c) Longitudinal sideband spectra at 560 E$_{\text{r}}$ after 3D cooling. Red sideband, carrier, and blue sideband transitions are highlighted with the red, grey, and blue line, respectively. (d) Blue sidebands at 560 E$_{\text{r}}$ under conditions of no clock-transition cooling (red), longitudinal sideband cooling alone (\emph{z} cooling, brown), radial cooling alone (\emph{xy} cooling, green), and 3D cooling (blue). In the green trace, we resolve transitions from different longitudinal lattice bands, whose frequency splitting is given by the trap anharmonicity.}
\end{figure}

Figure~1(c) shows the measured linewidth of the resulting velocity selection profiles, plotted as a function of the first 578-nm laser pulse duration. Data for the reference case of free-space atoms, measured by releasing the atoms from the lattice before the first pulse, closely follows the Fourier limited width (dashed line). Red diamonds show the case of atoms in a deep lattice (560 E$_{\text{r}}$). At pulse durations above 0.2 ms, spectral profiles are significantly broadened from modulation effects in the trap. Measured and simulated Rabi flopping from radial excitation also highlights the effect (see SM, \cite{Supp}). After introducing additional cooling, or utilizing lower lattice depths (which also exhibit lower initial temperatures), the measured linewidths lie closer to the Fourier limit. To summarize, radial atomic motion in the lattice limits the narrow velocity selectivity afforded by the 578-nm pulse, but the degradation is reduced as the atoms are more deeply cooled.

During radial cooling, both \emph{x} and \emph{y} 578-nm lasers are pulsed on at the same time, followed by a 1388-nm laser (1 mW, 20-$\mu$s pulse duration) to bring the population back to the ground state. The 578-nm laser $\pi$ pulses are tailored in duration and detuning to optimize the velocity selection, with pulses becoming longer and detuning smaller as the atoms get colder. The cooling pulse sequence, optimized for a lattice depth of 560 E$_{\text{r}}$, is shown in \cite{Supp}. We observe better results by cycling one pulse many times, then moving forward to the next pulse parameters, rather than repeating a sequence of each tailored pulse multiple times. As shown in Fig.~2(a), cooling reduces the linewidth of the radial spectrum from 118.4(12) kHz to 14.2(2) kHz, corresponding to a temperature decrease of nearly two orders of magnitude from 17.4(4) {\textmu}K to 250(10) nK. The cooled temperature represents a thermal energy equal to 0.4(1)\% of the trap depth. The cooled radial temperature lies below the recoil limit of 410 nK, given by cascaded spontaneous decay $^{3}$D$_{1}$-$^{3}$P$_{1}$-$^{1}$S$_{0}$, shown as the dashed line in Fig.~2(b). Nevertheless, the velocity selection linewidth measurements in Fig.~1(c) suggest that temperatures below 100 nK should be possible. As has been observed in Raman cooling, side lobes in the excitation spectrum may degrade the dark state, and Blackman pulses could offer lower temperatures at the cost of increased cooling time \cite{Raman1992}. Figure~2(b) shows the measured temperature versus cooling time, yielding a time constant of 13.8(10) ms. The possibility of fast cooling is beneficial to lattice clocks, to minimize the Dick effect \cite{Dick1987}. We also cooled samples at lattice depths of 48 E$_{\text{r}}$ and 115 E$_{\text{r}}$, reaching colder temperatures at 90(10) nK. We note that the pumping process at 1388 nm leads to $2.6\%$ decay to the long-lived $^{3}$P$_{2}$ state per cycle. With many repeated cooling cycles, we observe as much as 75$\%$ population loss at 560 E$_{\text{r}}$. The addition of another laser to optically pump the population out of $^{3}$P$_{2}$ could eliminate the loss.


\begin{figure}[t]
\includegraphics{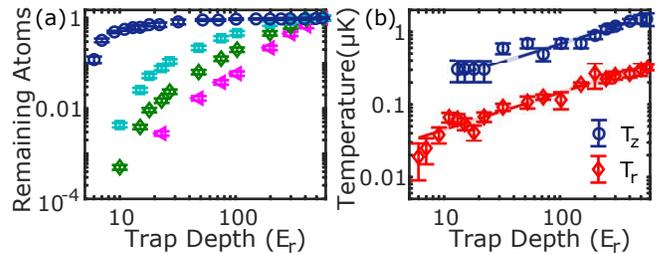}
\caption{\label{fig:3} (a) Remaining atomic fraction after adiabatically ramping to the indicated trap depth from 560 E$_{\text{r}}$ after 3D cooling at 560 E$_{\text{r}}$ (blue circle), after longitudinal sideband cooling alone (cyan square), or with no clock transition cooling at all (green diamond). Also included is the case of direct loading to a fixed lattice of indicated trap depth with no adiabatic ramp (magenta triangle). (b) Longitudinal temperature (T$_{\text{z}}$, blue circle) and radial temperature (T$_{\text{r}}$, red diamond) after adiabatically ramping to the indicated trap depth from 560 E$_{\text{r}}$ after 3D cooling at 560 E$_{\text{r}}$. The dashed lines are fit with the expected adiabatic scaling.}
\end{figure}

By combining pulsed radial cooling with longitudinal sideband cooling \cite{Lattice2017, Supp, Lattice2016}, atomic samples are cooled in all three dimensions. After cooling in one dimension, we typically observe residual heating in other dimensions. We therefore interleave sideband cooling with pulsed radial cooling. Figure~2(c) shows longitudinal sideband spectra at 560 E$_{\text{r}}$ after 3D cooling. The red sideband is virtually gone since the atomic population resides in the ground lattice band. In Fig.~2(d), we show blue-detuned sidebands at 560 E$_{\text{r}}$ under four different cooling situations. With no clock-transition cooling, the sideband exhibits a broad structure. After the application of pulsed radial cooling alone, distinct longitudinal lattice band transitions are well resolved, permitting measurement of the atomic distribution across the bands. On the other hand, after longitudinal sideband cooling alone, atoms occupy the longitudinal ground motional band, and the long tail of the sideband is due entirely to radial temperature. Finally, after 3D cooling, the width of the remaining sideband is dramatically narrowed, with virtually all population in the ground lattice band and cooled radially below the recoil limit.


Armed with efficient 3D cooling on the clock transition, we load large atom numbers into a deep magic-wavelength lattice and then adiabatically ramp to shallow depths. Blue circles in Fig.~3(a) show the remaining population at the final trap depth, normalized to the initial population at 560 E$_{\text{r}}$. The majority of the population is preserved at depths $\leq10$ E$_{\text{r}}$ $\approx$ k$_{\text{B}} \times $1 {\textmu}K. In absolute terms, we load thousands of atoms into 6 E$_{\text{r}}$ trap, which in this case is the lowest trap depth we can reach due to lattice tilt away from gravity. Figure~3(a) offers a comparison to cases with no cooling or longitudinal sideband cooling alone prior to the adiabatic ramp. Furthermore, the magenta triangles give the case of direct loading to a fixed lattice of indicated depth (with no adiabatic ramp). Ensuring that shallow lattices support high atom numbers is important for reaching excellent clock stability from quantum projection noise (QPN) \cite{QPN1993}. Using short duration $\pi$/2 excitation pulses, we measured shot-to-shot fluctuations in the atomic excitation to assess our detection signal-to-noise ratio (SNR). With atom numbers from $N=100$ to $N=2\times10^{4}$, we observe an SNR scaling as $1/\sqrt{N}$, as expected for QPN. For the 1 Hz spectral linewidth with which we typically operate the lattice clock, this corresponds to an atomic detection-limited clock stability of $1.4\times$10$^{-17}$/$\sqrt{\tau}$ for averaging time $\tau$ in seconds.

We measure atomic temperature spectroscopically after adiabatic ramping , with longitudinal sideband spectra for longitudinal temperatures \cite{Sb2009} and Doppler-broadened radial spectra for radial temperatures \cite{Supp}. As shown in Fig.~3(b), the results follow the expected adiabatic scaling 1/$\sqrt{\text{U}}$ (for trap depth $\text{U}$). At 6 E$_{\text{r}}$, the radial temperature is as low as 20 nK. For trap depths $\leq10$ E$_{\text{r}}$, no longitudinal temperature is plotted, since only the ground lattice band is trapped and motional sidebands are no longer present.          


We consider the immediate benefits of 3D cooling for lattice clock operation. In our systematic uncertainty evaluation of two Yb lattice clocks at the $1.4\times10^{-18}$ fractional frequency level \cite{Yb2018}, a dominant systematic uncertainty contributor stemmed from lattice light shifts. By operating with a lattice depth of 6 E$_{\text{r}}$ as shown here, together with an improved characterization of polarizability from magnetic dipole and electric quadrupole couplings, lattice light shifts uncertainty can be reduced to the $1\times10^{-19}$ level. With more precise measurement of the magic wavelength, even lower uncertainties are possible. Furthermore, based on measurements of lattice-induced $^{3}$P$_{0}$ quenching \cite{WillThesis, Jacob2022}, the quenching rate at 6 E$_{\text{r}}$ is more than an order of magnitude smaller than the spontaneous decay rate, yielding negligible impact on clock stability for interrogation times up to the clock state natural lifetime \cite{Supp}. Finally, ultracold atomic samples suppress p-wave inelastic losses that degrade spectroscopic contrast on the clock transition \cite{Srscatter2019}.

A potential drawback of shallow lattices is the increased intersite tunneling that can lead to motional frequency shifts during laser interrogation \cite{Sh2005}. A typical strategy to mitigate these effects is to use the Wannier-Stark lattice, which aligns the optical lattice along gravity. Gravity lifts the energy degeneracy between adjacent lattice sites, inducing atomic localization via periodic Bloch oscillations \cite{Bloch}. We observe prominent Bloch oscillations for the shallowest lattices used here. Figure~4(a) shows a longitudinal sideband spectrum at 12 E$_{\text{r}}$ lattice depth. We observe first-order Bloch oscillation sidebands at $\pm \Delta_{g}=mg\lambda_{l}/2\hbar \approx 1593$ Hz (where $g$ is the local gravitational acceleration) around the carrier at zero detuning, as well as Bloch sidebands around the blue motional sideband near 10 kHz (indicating combined motional excitation and tunneling). Figure~4(b) displays the Bloch sideband spectrum for a 6 E$_{\text{r}}$ lattice when excited by a 30 ms carrier $\pi$ pulse, yielding 91$\%$ carrier excitation and 10$\%$ excitation of the first-order Bloch sideband. In Fig.~4(c), we use measured excitation to deduce the relative Rabi frequencies $|\Omega_{0}/\Omega|^{2}$ and $|\Omega_{\pm1}/\Omega|^{2}$ as a function of trap depth, where $\Omega_{0}$, $\Omega_{\pm1}$, and $\Omega$ are the Rabi frequencies of the carrier transition, the first-order Bloch oscillation transition, and an atom in free space, respectively. Dashed lines give a theoretical calculation based on overlap integrals of the Wannier-Stark wave functions \cite{Supp} for atoms with finite radial temperature and accounting for radial gravitational sag due to slight lattice tilt with respect to gravity. 

\begin{figure}[t]
\includegraphics{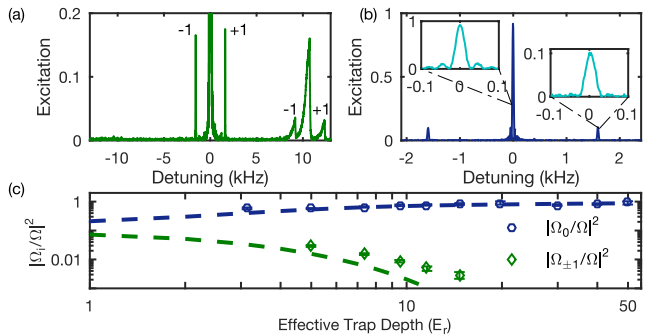}
\caption{\label{fig:4} (a) Longitudinal sideband and Bloch oscillation spectrum at 12 E$_{\text{r}}$. First-order Bloch oscillations ($\pm$1) are observed not only around the carrier transition but also around the blue sideband. (b) 30 ms $\pi$ pulse Rabi spectroscopy at 6 E$_{\text{r}}$. The insets are zoomed-in views of the carrier spectrum and the first-order Bloch oscillation spectrum. (c) Measured relative Rabi frequencies of the carrier and the first-order Bloch oscillation transition as a function of trap depth. Dashed lines are theoretical calculations.}
\end{figure}

In the presence of Bloch oscillations, tunneling leads to frequency broadening and shifts on the order of $\Omega_{\pm1}\Omega_{0}/\Delta_{g}$ \cite{Sh2005}. While the average tunneling shift depends on the relative phase of the atoms across different lattice sites, for our typical 560 ms Rabi spectroscopy, the maximum effect can only be $10^{-19}$ level. Therefore, these shallow lattices can support substantially improved clock accuracy of the future.

Very recent work in strontium \cite{Jun2022} highlights another potential complication of shallow lattices: s-wave atomic collision shifts mediated by tunneling. While more experimental investigation could be useful, we note that this effect is less relevant for $^{171}$Yb, where tunneling in a Wannier-Stark lattice is more strongly suppressed by atomic mass.

In conclusion, we demonstrate a pulsed cooling scheme achieving radial atomic temperature in a 1D lattice in the nK regime, below the recoil limit. Combined with longitudinal sideband cooling, we realize fast ultracold temperatures in all three dimensions, allowing for efficient transfer of atoms to a shallow lattice, where lattice light shifts and $^{3}$P$_{0}$ excited state quenching are strongly suppressed. Finally, we observe Bloch oscillations over a range of trap depths with tunneling shifts bounded at the low $10^{-19}$ level. This work paves the way for next-generation lattice clock uncertainty and stability, as well as enhanced control in quantum computation and simulation experiments \cite{Supp}.
\begin{acknowledgments}
We appreciate experimental assistance from R. Brown, C. Oates, and D. Nicolodi, as well as R. Brown and C. Oates for careful reading of the manuscript. This work was supported by NIST, ONR, and NSF QLCI Award OMA-2016244.
\end{acknowledgments}

\providecommand{\noopsort}[1]{}\providecommand{\singleletter}[1]{#1}%

\section{Supplemental Material: Sub-recoil clock-transition laser cooling enabling shallow optical lattice clocks}
\subsection{Radial Excitation and Velocity Selection on the Clock Transition}
While atoms reside in the Lamb-Dicke and well-resolved-sideband regimes along the longitudinal axis of our 1D optical lattice, this is not the case along the radial axes of the lattice. Consequently, radial excitation on the clock transition yields spectroscopic profiles that are Doppler-broadened. Figure~1 plots the measured time evolution of the excited state when driven by the 578-nm clock laser from the radial direction. In this case, laser power of approximately 5 mW is able to resonantly drive atoms near zero velocity with a Rabi frequency of $>$ 3 kHz. Excitation is shown for three different lattice depths. The excitation maximum around 0.1 ms occurs for a $\pi$ pulse area, with Rabi flopping that damps afterward due to inhomogeneous (thermal) dephasing. We typically operate near a $\pi$ pulse area for velocity selection in the pulsed radial cooling scheme. Excitation here is limited by the fraction of atoms resonantly driven within the overall velocity distribution. Note that as pulse time is extended up to and beyond the period of radial trap oscillation (tens of Hz), excitation increases with oscillations that are damped due to atomic inhomogeneity and trap anharmonicity. The high excitation ratio observed here (above $70 \%$) indicates that most of the velocity distribution is being excited, corresponding to poor velocity selection at these longer pulse times.

\begin{figure}[htbp]
\includegraphics{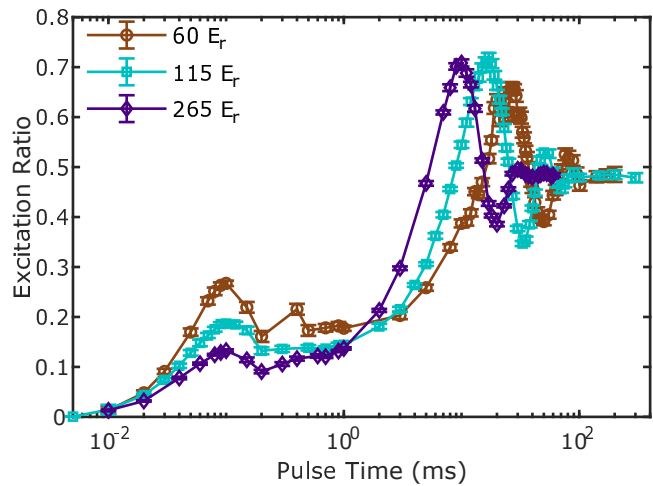}
\caption{\label{fig:1S} Measured excited-state evolution with time when driven by the 578-nm clock laser from the radial direction at different trap depths.}
\end{figure}

As noted in the main text, for typical conditions we observe that $\pi$ pulses with pulse times more than several hundred {\textmu}s can lead to velocity selection profiles with broadened linewidths. In contrast to the case of shorter pulses that better sample the instantaneous atomic velocity in the trap, radial atomic motion is not negligible on the longer excitation pulse time scale. Figure~2(a) shows the measured spectral linewidths of the velocity selection profiles as a function of the velocity selection pulse time, for a $\pi$ pulse area. Both red and blue points indicate experimental data measured for atoms in an optical lattice depth of 560 E$_{\text{r}}$, with the blue points for atoms that were first cooled on the clock transition in three dimensions. Dotted lines are computed curves based on a 1D Monte Carlo simulation of 1000 atoms undergoing harmonic motion at the radial trap frequency. The amplitude of motion for each atom was taken from an initial distribution given by the radial temperature. The simulation solves the equations of motion for a driven two-level atom in small time steps during which the atomic velocity is taken as constant. After each time step, atomic excitation and motion are updated.  While agreements with experiments are reasonable, we anticipate better quantitative agreement by extending the simulation to three dimensions and more carefully including residual trap anharmonicity.

\begin{figure}[htbp]
\includegraphics{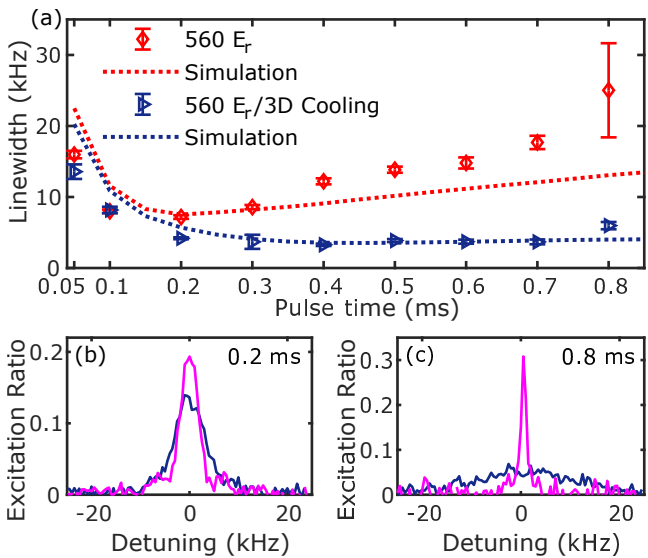}
\caption{\label{fig:2S} (a) Measured spectral linewidths of the velocity selection profiles and the simulation data as a function of the velocity selection pulse time at 560 E$_{\text{r}}$ without or with  3D cooling. Note that the radial trap frequency at 560 E$_{\text{r}}$ is 96 Hz. (b) Measured velocity selection spectra with atoms in the 560 E$_{\text{r}}$ lattice (blue) and free space (magenta) for the velocity selection pulse duration of 0.2 ms. (c) Measured velocity selection spectra with atoms in the 560 E$_{\text{r}}$ lattice (blue) and free space (magenta) for the velocity selection pulse duration of 0.8 ms. }
\end{figure}

Figures~2(b) and 2(c) show individual velocity selection spectra that were experimentally measured with 0.2-ms and 0.8-ms velocity selection pulse durations. Blue curves give the results for lattice trapped atoms (560 E$_{\text{r}}$), whereas magenta curves offer a comparison to the case of atoms in free space. Broadening due to the radial motion in the lattice is prominent for the longer pulse time. Note that in the main text Fig.~1(c), velocity selection profile linewidths are shown for atoms both in lattice and in free space. In lattice, the measured spectral linewidth corresponds to the full-width at half-maximum (FWHM) of a Gaussian fit. For the case of free space, the FWHM of a sinc-squared function fit is used.          

\subsection{Radial Cooling Scheme}
Figure~3(a) shows the optimal pulsed radial cooling time sequence for a trap depth of 560 E$_{\text{r}}$. Clock laser pulses are tailored in intensity, duration, and frequency detuning to optimize the velocity selection. Acousto-optic modulators (AOMs) are used for defining the pulse duration, varying the laser intensity, and changing the frequency detuning. The AOMs' turn-on and turn-off time is measured to be less than 10 {\textmu}s. Similar pulse sequences are also realized at 115 E$_{\text{r}}$ and 48 E$_{\text{r}}$ with a last-pulse duration of 0.7 ms, which are shown in Fig.~3 (b) and (c).

\begin{figure}[htbp]
\includegraphics{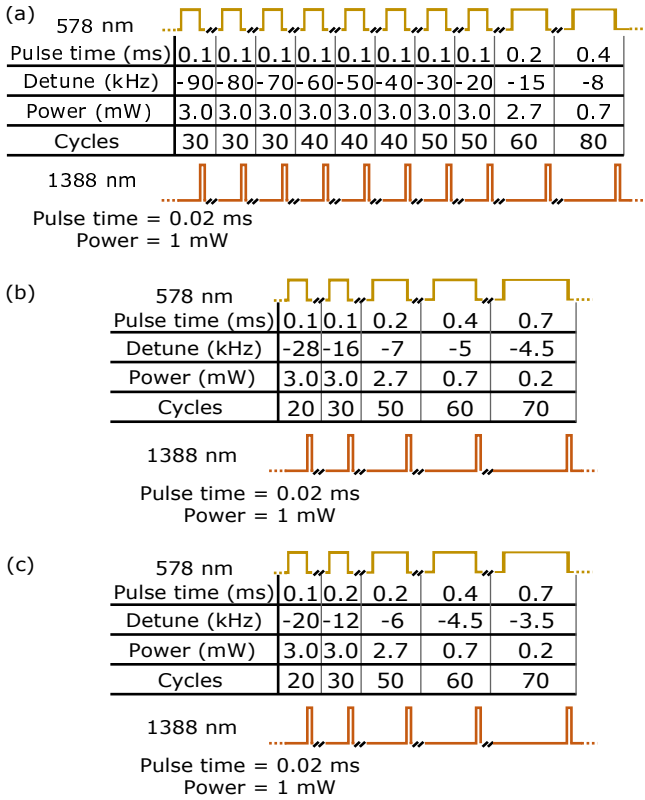}
\caption{\label{fig:3S} (a) Optimal pulsed radial cooling time sequence at 560 E$_{\text{r}}$. For a chosen power and detuning, two pulses, one 578-nm pulse followed by a short 1388-nm pump pulse, are interleaved and repeated for the indicated number of cycles. The 578-nm pulses become longer as the detuning is closer to zero. (b) Optimal pulsed radial cooling time sequence at 115 E$_{\text{r}}$. (c) Optimal pulsed radial cooling time sequence at 48 E$_{\text{r}}$.}
\end{figure}

\subsection{Radial Temperature}
Radial temperature in the 1D lattice is extracted from fitting the Doppler-broadened radial spectra. While we observe Gaussian-shaped spectra prior to radial cooling, the spectra after cooling can be somewhat more complex. Figure~4 shows a cooled radial spectrum at 560 E$_{\text{r}}$ with different line shape fits. Due to the excellent fitting of the Voigt profile over a wide range of experimental conditions, and for the sake of consistency, we opt to employ this fit and its FWHM for all measurements of radial temperature throughout this work. For comparison, note that in Fig.~4, the extracted temperatures from the FWHM of the Lorentz fit, the Gaussian fit, and the Voigt fit are 216(10) nK, 203(10) nK, and 250(10) nK, with fitting R-value of 0.9924, 0.9885, and 0.9968, respectively.  

\begin{figure}[t]
\includegraphics{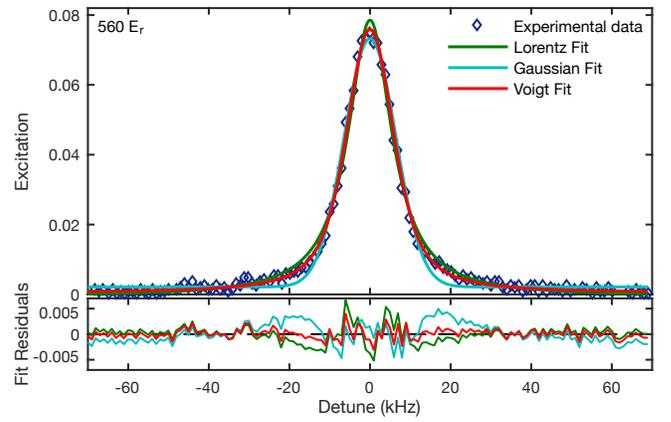}
\caption{\label{fig:4S} Cooled radial spectrum at 560 E$_{\text{r}}$ with different line shape fits. Fit residuals are also shown.}
\end{figure}

\subsection{Longitudinal Sideband Cooling}
Longitudinal sideband cooling is realized by tuning the clock laser frequency to the (first-order) red motional sideband and coherently driving atoms from higher to lower motional states. These atoms are then optically pumped from the excited clock state with 1388-nm laser light back to the ground state. When atoms reach the ground lattice band, no excitation occurs. The clock laser and optical pumping laser can be turned on sequentially or simultaneously, and we found the latter to be somewhat faster cooling. In this case, the 1388-nm laser power is decreased to 1 {\textmu}W to avoid light shifts on the $^3$P$_0$ state. For best cooling efficiency, the clock laser is tuned to the frequency where maximum excitation in the red sideband is observed. No added modulation on the 578-nm excitation laser is required. The cooling time normally is from 10 ms to 100 ms depending on trap depths. Generally, higher trap depths need longer cooling time with the same 578-nm clock laser power.

\subsection{3D Cooling Scheme}
Combining longitudinal sideband cooling and pulsed radial cooling, the atomic sample is cooled in all three dimensions. In order to mitigate residual heating between dimensions, we interleave the longitudinal sideband cooling and the pulsed radial cooling process. A specific interleaved time sequence for a trap depth of 560 E$_{\text{r}}$ is shown in Fig.~5. We start with longitudinal sideband cooling and also end with it. Longitudinal sideband and pulsed radial cooling time are gradually decreased after each interleaved step. In the interleaved process, the clock laser frequency is stepped in the longitudinal sideband cooling scheme to optimize the cooling efficiency.

\begin{figure}[htbp]
\includegraphics{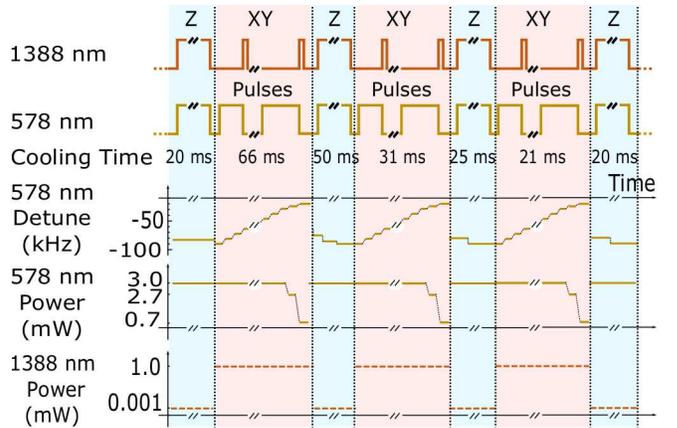}
\caption{\label{fig:5S} Interleaved 3D cooling scheme at 560 E$_{\text{r}}$. The 1388-nm laser power is shown with the red dashed line.}
\end{figure}

\subsection{Wannier-Stark Wave Function}
The theoretical calculation in the main text Fig.~4(c) is based on overlap integrals of the Wannier-Stark wave functions. For the ground motional band (along the lattice axis), the normalized Wannier-Stark wave functions are given by
\begin{align*}
\psi_j(z)=
{}&\frac{1}{\sqrt{\pi}}\int_{0^+}^{1^-}
\left[
\cos\chi_{\nu,j}
\,ce_\nu\left(\frac{2\pi z}{\lambda_l},-\frac{U}{4E_r}\right)
\right.\\&\left.
+
\sin\chi_{\nu,j}
\,se_\nu\left(\frac{2\pi z}{\lambda_l},-\frac{U}{4E_r}\right)
\right]d\nu,
\end{align*}
where $j$ is an integer lattice site index and
\begin{gather*}
\chi_{\nu,j}=\pi\frac{2E_r}{mg\lambda_l}\left(\nu\gamma_1-\gamma_\nu\right)+\pi j\nu,
\\
\gamma_\nu=\int_0^\nu a_{\nu^\prime}\left(-\frac{U}{4E_r}\right)d\nu^\prime.
\end{gather*}
Here $ce_\nu(x,q)$ and $se_\nu(x,q)$ are Mathieu functions, while $a_\nu(q)$ is the Mathieu characteristic value. The Mathieu functions $ce_\nu(x,q)$ and $se_\nu(x,q)$ are real-valued and are even and odd, respectively, with respect to the argument $x$. For rational $\nu$, these functions are also periodic in the argument $x$ and satisfy the normalization
\begin{gather*}
\int_L\left[ce_\nu\left(x,q\right)\right]^2dx=\int_L\left[se_\nu\left(x,q\right)\right]^2dx=L/2,
\end{gather*}
where the integrals extend over a segment of length $L$ that is any multiple of a period. For irrational $\nu$, we assume the same normalization with $L\rightarrow\infty$. We note that Landau-Zener tunneling is neglected here, which we estimate to be small.

\begin{figure}[t]
\includegraphics{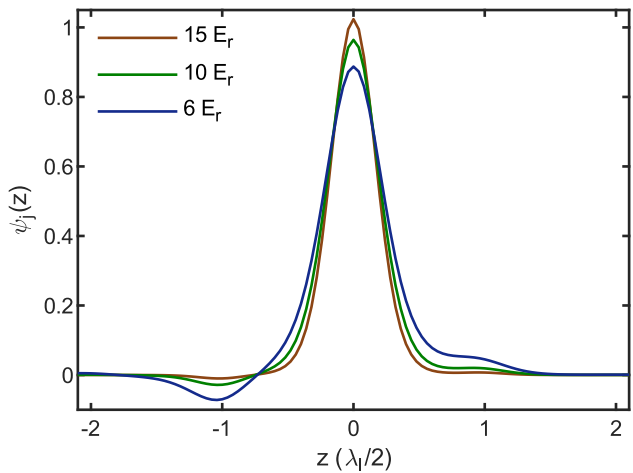}
\caption{\label{fig:6S} Ground motional band Wannier-Stark wave functions at 6 E$_{\text{r}}$, 10 E$_{\text{r}}$, and 15 E$_{\text{r}}$.}
\end{figure}
Figure~6 shows the ground motional band ($n = 0$) Wannier-Stark wave functions for 6 E$_{\text{r}}$, 10 E$_{\text{r}}$, and 15 E$_{\text{r}}$. The Rabi frequency of optically-driven Bloch oscillations are calculated as  $\Omega_{j} = \Omega \langle \psi_0(z)|e^{ik_{s}z}|\psi_j(z) \rangle $, where $\Omega$ is the Rabi frequency of an atom in free space and $k_{s}$ is the wave vector of the clock laser.

\begin{table*}[t]
\caption{\label{tab:Transitions} Transitions used in this cooling technique for different species.}
\begin{tabular}{|c|c|c|c|c|}
\hline
\textbf{Parameters} & \textbf{$^{171}$Yb} & \textbf{$^{87}$Sr} & \textbf{$^{199}$Hg} & \textbf{$^{111}$Cd}\\
\hline
Cooling transition & ${^1\!S_0}\leftrightarrow{^3\!P_0}$ & ${^1\!S_0}\leftrightarrow{^3\!P_0}$ & ${^1\!S_0}\leftrightarrow{^3\!P_0}$ & ${^1\!S_0}\leftrightarrow{^3\!P_0}$ \\
\hline
Wavelength [nm] & 578 & 698 &  266 & 332  \\
\hline
Excited-state linewidth [mHz] & $\sim$8 & $\sim$1 & $\sim$100 & $\sim$7 \\
\hline
Magic wavelength lattice [nm]  & 759 & 813 & 363 & 420 \\
\hline
Typical repump transition  &  ${^3\!D_1}\leftrightarrow{^3\!P_0}$  & \begin{tabular}{c} ${^3\!S_1}\leftrightarrow{^3\!P_0}$ \\ ${^3\!S_1}\leftrightarrow{^3\!P_2}$ \end{tabular} &  \begin{tabular}{c} ${^3\!S_1}\leftrightarrow{^3\!P_0}$ \\ ${^3\!S_1}\leftrightarrow{^3\!P_2}$ \end{tabular} & \begin{tabular}{c} ${^3\!S_1}\leftrightarrow{^3\!P_0}$ \\ ${^3\!S_1}\leftrightarrow{^3\!P_2}$ \end{tabular}  \\
\hline
Wavelength [nm] &  1388  & \begin{tabular}{c} 679 \\ 707 \end{tabular}  &   \begin{tabular}{c} 405 \\ 546 \end{tabular} &  \begin{tabular}{c} 468 \\ 509 \end{tabular} \\
\hline
Linewidth [MHz]      & 0.48  & \begin{tabular}{c} 1.4 \\ 6.7 \end{tabular} &  \begin{tabular}{c} 3.3 \\ 7.8 \end{tabular} & \begin{tabular}{c} 2.1 \\ 8.9 \end{tabular} \\
\hline
\end{tabular}
\end{table*}

\subsection{Cooling in other atomic species}
This clock transition cooling can be applied to other atomic species. Here we list some of the relevant information for other prospective species in Table~1.

\end{document}